\documentclass[10pt]{iopart}

\usepackage{color}
\usepackage{graphicx}
\usepackage[utf8]{inputenc}

\expandafter\let\csname equation*\endcsname\relax
\expandafter\let\csname endequation*\endcsname\relax

\usepackage{mathtools}
\usepackage{amsfonts}
\usepackage{amssymb}

\newcommand{\tikzFigure}[1]{\includegraphics{tikz-cache/#1.pdf}}

\newcommand{\eqnref}[1]{(\ref{#1})}
\newcommand{\figref}[1]{figure~\ref{#1}}

\newcommand{\direction}[1]{\ensuremath{\hat{\bf e}_{#1}}}
\newcommand{\directionX}{\direction{x}}
\newcommand{\directionY}{\direction{y}}
\newcommand{\directionZ}{\direction{z}}
\newcommand{\directionB}{\direction{\bf B}}

\begin{document}

\title[Targeted fractionation of magnetic clusters via shear
  flow]%
{%
	Controlled self-aggregation of polymer-based nanoparticles
	employing shear flow and magnetic fields
}

\author{David Toneian$^{1,2}$, Christos N. Likos$^2$, and Gerhard Kahl$^1$}

\address{$^1$Institute for Theoretical Physics, TU Wien, Wiedner
  Hauptstra{\ss}e 8-10, A-1040 Vienna, Austria} 
\address{$^2$Faculty of Physics, University of Vienna, Boltzmanngasse
  5, A-1090 Vienna, Austria}

\ead{david@toneian.com}

\begin{abstract}
Star polymers with magnetically functionalized end groups are
presented as a novel polymeric system whose morphology,
self-aggregation, and orientation
can easily be tuned by exposing these macromolecules
simultaneously to an external
magnetic field and to shear forces. Our investigations are based on a
specialized simulation technique which faithfully takes into account
the hydrodynamic interactions of the surrounding, Newtonian solvent.
We find that the combination of magnetic field
(including both strength and direction)
and shear rate controls the mean number of magnetic clusters,
which in turn is largely responsible for the static and dynamic behavior.
While some properties are similar to comparable non-magnetic
star polymers, others exhibit novel phenomena;
examples of the latter include
the breakup and reorganization of the clusters
beyond a critical shear rate,
and
a strong dependence of the
efficiency with which shear rate is translated into whole-body rotations
on the direction of the magnetic field.
\end{abstract}

\maketitle
\ioptwocol

\section{Introduction}
Star polymers, a family of
macromolecules where $f$ polymeric arms (each consisting of $n_{\textrm{A}}$
monomers) are tethered to a central, colloidal particle, have received
a rapidly increasing share of interest within the soft matter
community during the past years (see e.g. \cite{Likos:2006,Vlassopoulos:2014}).
The reason for
their popularity rests both upon the tunability of their architecture
via variations of $f$ and/or $n_{\rm A}$ as well as upon the
possibility to functionalize star polymers by selectively designing
the polymeric arms \cite{Garlea:2017}.
This functionalization can, for instance, be
realized by tethering block copolymers to the central colloid,
leading to so-called telechelic star polymers \cite{Garlea:2017};
alternatively, 
as recently put forward in \cite{Blaak:2018},
one can attach (super-para-)magnetic particles as terminal monomeric units
onto each of the arms.
This latter manner of functionalization is particularly attractive in that
it allows for well-controlled and practically instantaneous
tuning of the interaction, and hence of the system properties,
via the external magnetic field,
so that one does not have to rely on slow and inaccurate changes
in temperature.
In addition, it introduces strong anisotropy of the interactions
between the endgroups, modifying thereby the morphology of the
terminal aggregates from spherical into linear ones \cite{Blaak:2018}.

Star polymers show in their different architectures a broad range of
interesting physical equilibrium properties;
Examples include
(i) the ability to cover in their
single-molecule properties -- by tuning their functionality $f$ -- the range of
ultrasoft to spherical, essentially hard colloidal particles
\cite{Likos:2006,Vlassopoulos:2014},
(ii)
association, where telechelic star polymers form self-assembled,
reconfigurable, soft patchy colloids, which then further
self-organize at a supramolecular level into a variety of micellar
or network-forming structures \cite{Garlea:2017,Bianchi:2015},
and
(iii) the ability of the above-mentioned magnetically functionalized star polymers
to form under equilibrium conditions clusters of particles
(``valences''), whose number and size depend on $f$, $n_{\rm A}$, and
the strength and the orientation of the external magnetic field ${\bf B}$.
The wealth of emerging
scenarios (in terms of valence and molecular shape) has been
thoroughly discussed in \cite{Blaak:2018}.
In addition to equilibrium situations, conventional star
polymers exhibit a variety of intriguing properties
in a stationary, non-equilibrium setup, as demonstrated
in the investigations by Ripoll {\it et al.}~\cite{Ripoll:2006}, who have
exposed these macromolecules to shear forces by faithfully including
hydrodynamic interactions: depending on the values of $f$ and $n_{\rm
  A}$, the particles show, upon increasing the shear rate $\dot \gamma$,
strong deformations and distinctively different types of motion.

In this contribution, we extend these non-equilibrium simulations to
the aforementioned magnetically functionalized star polymers and
expose these particles both to shear forces as well as to an external
magnetic field ${\bf B}$, considering three orientations
of the latter relative to
the shear flow direction (\directionX),
the shear gradient direction (\directionY),
and the vorticity direction (\directionZ).
As compared to the related investigations on
conventional star polymers \cite{Ripoll:2006}, we face here an entirely
new situation due to the emergence of patches which can or cannot be
broken up under the influence of external fields,
their stability being governed by an interplay between shear rate,
magnetic field strength, and relative orientation
of ${\bf B}$ to the shear-cell geometry.

Employing the multi-particle collision dynamics (MPCD) technique
\cite{Gompper:2009}, which incorporates hydrodynamic interactions,
we provide evidence that conformational properties, such as the
number, the size, and the location of the magnetic clusters, the shape
of the macromolecule, or its flexibility can easily and accurately
(but not necessarily independently from one another) be triggered via
suitable combinations of the two above-mentioned external fields. With
this contribution we thus introduce magnetically functionalized star
polymers as a novel system of very flexible
particles featuring specific numbers of self-associating aggregates
with
versatile and easily addressable conformational properties.

To the best of our knowledge,
magnetically functionalized star polymers have not been synthesized
in experiment,
but realization of magnetic nanoparticles
\cite{Wang:2011},
their successful chemical coating and linkage
\cite{Zhou:2009},
and a rich history of the study of other types of star polymers
\cite{Hadjichristidis:2012}
make the synthesis of magnetically functionalized star polymers
feasible and render them, as we hope to show in the following,
interesting candidates for future experiments.

\begin{figure*}
\centering
(a)
\includegraphics[width=0.2\linewidth]{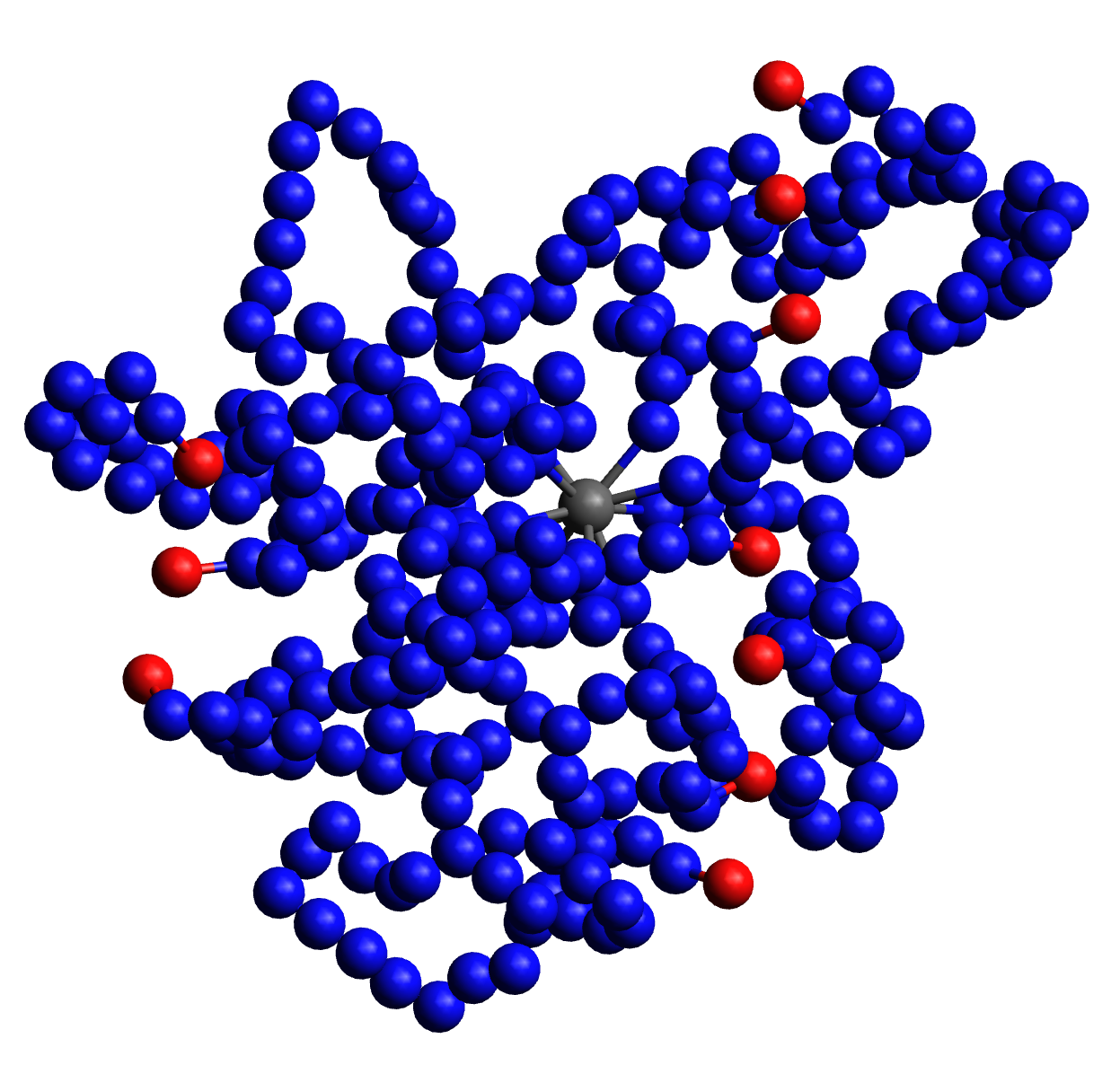}
(b)~~
\resizebox*{12.cm}{!}{\includegraphics{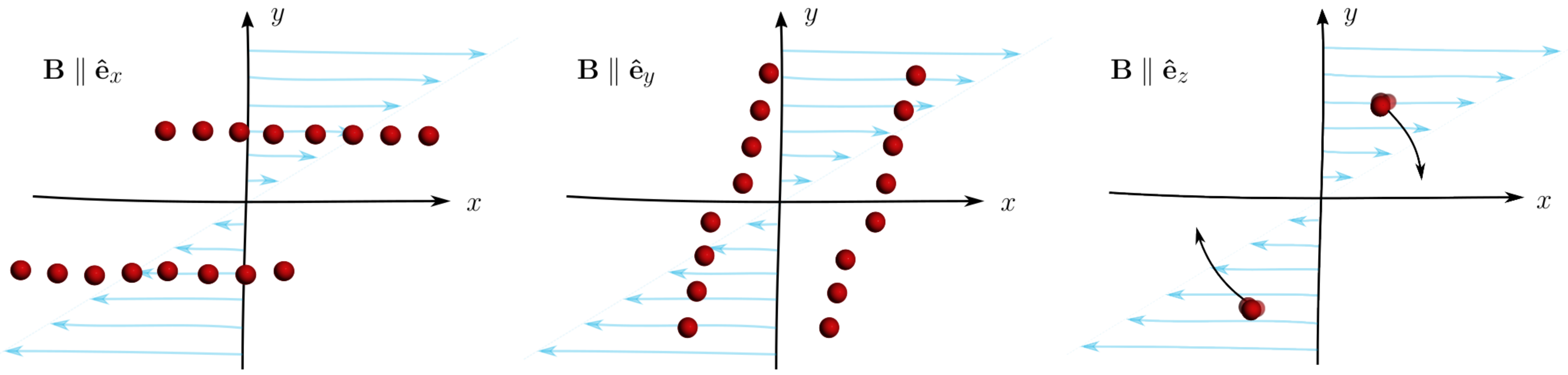}}
\caption{(color online)
  (a): Simulation snapshot, showing the star's core (gray), arm monomers (blue),
  and magnetic monomers (red).
  (b): Schematic representation of our simulation
  setup: a magnetically functionalized star polymer is exposed to
  shear flow (as specified in the text) and to an external magnetic
  field ${\bf B}$, pointing in independent experiments along the Cartesian
  axes. Blue: velocity profile of the flow, red: schematic
  representation of the super-paramagnetic end monomers, forming two
  magnetic clusters (suppressing the arm monomers and the central core bead,
  which would be situated approximately in the origin in this sketch).
  In the three panels, the magnetic field ${\bf B}$ points along the flow, gradient,
  or vorticity direction (from left to right).
  }
\label{fig:schematic}
\end{figure*}

\section{Model and Methods}
In our investigations,
we employ a bead-spring model for the
magnetically functionalized star polymers:
$f$ linear polymer arms are attached to a
core particle (index 'C'), each of them containing $n_{\rm A}$ arm
particles (index 'A'); to the end of each arm, a super-paramagnetic
particle (index 'M') is attached. The steric interactions of all these
spherical monomeric units have two concentric interaction ranges: an
inner, impenetrable part with diameter $D_\alpha$ and an outer, soft part
with range $\sigma_\alpha$, with $\alpha$ = C, A, or M. The masses of
all types of monomers are assumed to be equal
in order to avoid introducing features dependent on
specific mass asymmetries.

Any pair of monomers, separated by a distance $r$, interact via a
modified Weeks-Chandler-Andersen (WCA) potential $V_{\rm WCA}(r)$ \cite{Weeks:1971},
given by
\begin{eqnarray*}   \label{WCA}
V_{\rm WCA}(r) 
=
\begin{cases}
V_0 (r)
& {\rm if } ~~~r - D_{\alpha\beta} \le 2^{1/6} \sigma_{\alpha\beta},
\\
0 & \textrm{else}
\end{cases},
\\
V_0 (r)
=
4 \epsilon_{\alpha\beta}
\left[
	\left( \frac{ \sigma_{\alpha\beta} }{ r - D_{\alpha\beta} } \right)^{12}
	-
	\left( \frac{ \sigma_{\alpha\beta} }{ r - D_{\alpha\beta} } \right)^{6}
	+
	\frac{1}{4}
\right]
,
\end{eqnarray*}
with $D_{\alpha \beta} = (D_\alpha + D_\beta)/2$, $\sigma_{\alpha
  \beta} = (\sigma_\alpha + \sigma_\beta)/2$, and $\epsilon_{\alpha
  \beta} = \sqrt{ \epsilon_\alpha \epsilon_\beta}$,
to be set to specific values in what follows.

Spring bonds between (i) the core monomer and the first arm
monomer, (ii) adjacent arm monomers, and (iii) the last arm
monomer and the functionalized monomer are modeled via the generalized
finitely extensible non-linear elastic (FENE) potential \cite{Bird:1987a,Warner:1972},
specified via
\begin{equation}
V_{\rm FENE}(r)  =
- \frac{1}{2} K_{\alpha\beta} R_{\alpha\beta}^2
\ln \left[1 - \left( \frac{ r - l_{\alpha\beta} }{R_{\alpha\beta} } \right)^2
\right] ;
\end{equation}
here, $K_{\alpha \beta}$ specifies the interaction strength, $l_{\alpha
  \beta}$ is the equilibrium bond length between monomers $\alpha$ and
$\beta$, and $R_{\alpha \beta}$ is the maximum deviation from
$l_{\alpha \beta}$.
In addition, the magnetic monomers interact via the standard
dipole-dipole interaction, i.e.,
\begin{equation} \label{dipolar}
V_{\rm M} ({\bf r}) = 
- \frac{\mu_0}{4 \pi r^3}
\left[ 3 \left( {\bf m}_1 \cdot \hat {\bf r} \right) 
\left( {\bf m}_2 \cdot \hat {\bf r} \right)
- {\bf m}_1 \cdot {\bf m}_2 \right] ;
\end{equation}
with
${\bf m}_1$ and ${\bf m}_2$ being the dipolar moments of two interacting
particles which are separated by a vector ${\bf r}$ (with $r = |{\bf
  r}|$ and $\hat {\bf r} = {\bf r}/r$). The dipole moments are assumed
to be equal in magnitude (i.e., $|{\bf m}_1| = |{\bf m}_2| = m$), and $\mu_0$ is
the vacuum permeability.

For reasons of simplicity, we assume that the moments of the
super-paramagnetic particles are always perfectly aligned with the
external, spatially homogeneous magnetic field, ${\bf B} = B \directionB$.
With all this in mind, the expression (\ref{dipolar}) for
$V_{\rm M}({\bf r})$ simplifies to
\begin{equation}
V_{\rm M} ({\bf r})
=
- \frac{\mu_0 m^2}{4 \pi r^3} \left[ 3 \left( \directionB \cdot \hat {\bf r} \right)^2 - 1 \right]
    .
\end{equation}
We introduce the dimensionless magnetic parameter $\lambda = \mu_0 m^2/(4
\pi \sigma^3 \epsilon)$, with the length and energy scales $\sigma$ and $\epsilon$
defined below.
$\lambda$ represents the relative strength of the magnetic interaction compared
with the other potentials, as well as the thermal and hydrodynamic interactions.
Assuming, due to the super-paramagnetism, that $ m \propto B $,
one can consider $ \lambda \propto B^2 $ a measure of the magnetic
field strength,
and thus view dependencies on $\lambda$ and $\directionB$ as
dependencies on the external ${\bf B}$-field in (computer) experiments.

In an effort to reduce the large number of system parameters, we have
used the following set of WCA-parameters which mimic a simple, yet
reasonable model of a magnetically functionalized star polymer:
$$
\epsilon_\alpha = k_{\rm B} T = \epsilon ~~~ {\rm with ~~~} \alpha = {\rm C, ~A, ~or~ M,}
$$
$$
D_{\rm C} = 2a, ~~~~~D_{\rm A} = 0, ~~~~~D_{\rm M} = a
,
~~~~~~{\rm and}
$$
$$
\sigma_{\rm C} = \sigma_{\rm A} = \sigma_{\rm M} = a = \sigma.
$$
Here,
$T$ is the temperature, $k_{\rm B}$ is Boltzmann's constant, and $a$
is the MPCD length unit, to be specified below%
\footnote{
	$\sigma = a$ has been chosen in order to achieve, on average,
	a spatial separation of two monomers sufficient to place them in
	different MPCD collision cells.
}.
For the FENE parameters we use $K_{\alpha \beta} = 30 \epsilon_{\alpha
  \beta} \sigma_{\alpha \beta}^{-2}$ and
$$
l_{\alpha \beta} = D_{\alpha \beta} ,~~~~ 
R_{\alpha \beta} = 1.5 \sigma_{\alpha \beta}, ~~~~ {\rm with ~~} 
 \alpha = {\rm C, ~A, ~or~ M}  .
$$

To quantify the shape of the star polymer under arbitrary external conditions,
we employ the radius of gyration tensor, ${\cal S}$, with
elements $S_{\mu \nu}$ ($\mu, \nu = 1, 2, 3$)
\begin{equation} \label{gyration_tensor}
S_{\mu \nu} = \frac{1}{N}
\sum_{i=1}^N r_\mu^{i} r_\nu^{i} ;
\end{equation}
$r_\mu^{i}$ is the $\mu$-th component of the Cartesian position
vector of particle $i$ with respect to the molecule's center of mass frame;
$N = 1 + f (n_{\rm A} + 1)$ is the total number of monomers.
From the eigenvalues of this tensor, termed $\Lambda^2_\alpha$
($\alpha = 1, 2, 3$), and assuming, without loss of generality,
that $\Lambda^2_1 \le \Lambda^2_2 \le \Lambda^2_3$, one can calculate
the acylindricity $c$, the asphericity $b$, the radius of gyration
$R_{\rm g}$, and the relative shape anisotropy $\kappa^2$ of the
macromolecule \cite{Theodorou:1985}:
\begin{eqnarray}   \label{shape_parameters}
c & = & \Lambda_2^2 - \Lambda_1^2 
~~~~~
b = \Lambda_3^2 - \frac{1}{2} \left( \Lambda_1^2 + \Lambda_2^2 \right)
\\ \nonumber
R_{\rm g} & = & \sqrt{\Lambda_1^2 + \Lambda_2^2 + \Lambda_3^2}
~~~~~
\kappa^2 = \frac{1}{R_{\rm g}^4}\left( b^2 + \frac{3}{4} c^2 \right) .
\end{eqnarray}
Typical configurations of functionalized star polymers under {\it
  equilibrium} conditions, for different arm lengths (i.e., different
values of $n_{\rm A}$) and different values of $\lambda$ under the
influence of some external magnetic field are shown in figure 1 of
\cite{Blaak:2018};
a key observation is the emergence of columns
of endgroup-monomers extending parallel to $\directionB$.

To shed light on the tunability of these particles
under {\it non-equilibrium} conditions, we have exposed in this
contribution a single functionalized star polymer to shear forces,
assuming the flow direction, the velocity gradient direction, and the
vorticity direction along the $x$-, $y$, and $z$-axes, respectively;
the strength of the flow is measured by the shear rate $\dot
\gamma$. In addition, we have applied an external magnetic field, ${\bf
  B}$, which we have assumed in distinct computer experiments
to be oriented along each of the Cartesian axes; see
\figref{fig:schematic} for a schematic representation.

To avoid a scan of the high-dimensional parameter space, we have
restricted ourselves to the case of star polymers with a functionality
$f = 10$, each arm being formed by $n_{\rm A} =30$ monomers --
a situation which is computationally very tractable and still exhibits
rich physics and phenomenology already in the equilibrium case \cite{Blaak:2018}.
For the
reduced magnetic interaction strength $\lambda$ two values have been
assumed, namely $\lambda = 100$ and $\lambda = 200$. From the diagrams
of states (as they are shown in \cite{Blaak:2018}) we know that
for this set of parameters, star polymers form under equilibrium
conditions two to three magnetic column-shaped clusters;
these are assemblies of
interacting magnetic end-monomers, aligned along the external
magnetic field, where two magnetic beads are
considered to be part of the same cluster if their interparticle
distance is at most $2.5 a$.

In the Multi-Particle Collision Dynamics (MPCD) technique, the
macromolecule is surrounded by microscopic fluid particles of mass
$m_{\rm f}$ which are considered point particles; their positions
and momenta are not constrained to a lattice (for details
cf. \cite{Gompper:2009}). In this simulation technique, two steps are
carried out alternately: (i) in the \textit{streaming step}, the point
particles move ballistically for a time $\Delta t$, such that ${\bf
  r}_i(t+\Delta t) = {\bf r}_i(t) + {\bf v}_i(t) \Delta t $, ${\bf
  r}_i(t)$ and ${\bf v}_i(t)$ being the position and velocity of
particle $i$, respectively. (ii) In the \textit{collision step},
interaction takes place: in the variant
of MPCD that we have employed in this contribution,
Stochastic Rotation Dynamics,
the point particles
are sorted, according to their instantaneous positions ${\bf r}_i(t)$,
into \textit{collision cells}, i.e.\ cubic boxes of side length $a$,
which tesselate the simulation volume. Then,
for each collision cell $k$, one transforms the velocities of all
particles $i$ in that cell according to the rule ${\bf v}_i(t) \mapsto
{\bf V}_k(t) + {\cal R}(k, t, \alpha) \left[ {\bf v}_i(t) - {\bf
    V}_k(t) \right]$, where ${\bf
  V}_k(t) = ( \sum_{i\in\textrm{cell}_k}m_i {\bf
  v}_i(t)/\sum_{i\in\textrm{cell}_k}m_i )$
is the center-of-mass
velocity of collision cell $k$, $m_i$ is the mass of particle $i$, and
${\cal R}(k, t, \alpha)$ is a rotation matrix about a randomly chosen
axis and fixed angle $\alpha$, with independent choices for each
collision cell $k$ and time $t$. In order to suspend the star polymer
in this MPCD fluid, its beads are treated like fluid particles, except
that their masses are $m_{\textrm{b}} = 5 m_{\textrm{f}}$, and --
instead of ballistic streaming -- the intra-star forces are integrated
in five consecutive iterations of a velocity-Verlet algorithm \cite{Frenkel:2002},
each with timestep $\Delta t/5$.

Simulations were initialized with equilibrium configurations of the
stars and a random fluid configuration; representative data was taken
only after an equilibration period to avoid correlations with the
initial state.  The simulation volume was chosen to be cubic and of
side length $30 a$.  Lees-Edwards boundary conditions \cite{Lees:1972}
were employed to enforce a shear flow. Units are chosen such that
$a=1$, $m_{\textrm{f}} = 1$, and $k_{\textrm{B}} T = 1$, the temperature $T$ being
enforced via the Maxwell-Boltzmann scaling thermostat
\cite{Huang:2010}. The pure fluid's mass density was set to $\varrho = 10$, such
that in total $10 \times 30^3 = 270 \times 10^3$ MPCD fluid particles were
simulated.
The rotation angle $\alpha$ was set to $2.27$ radians,
corresponding to approximately $130$ degrees.
The {\it OpenMPCD} simulation package used can be found at \cite{OpenMPCD}.

\begin{figure*}
\centering
\begin{tabular}{rl}
\tikzFigure{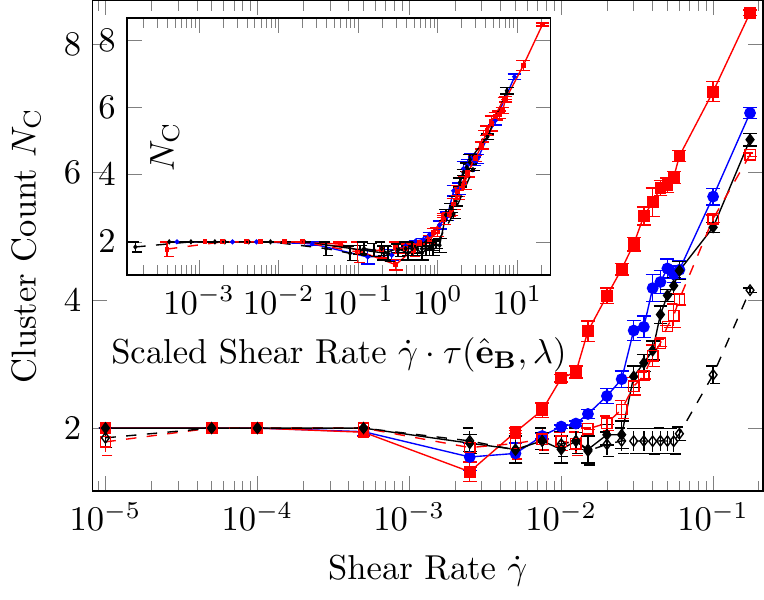}
&
\tikzFigure{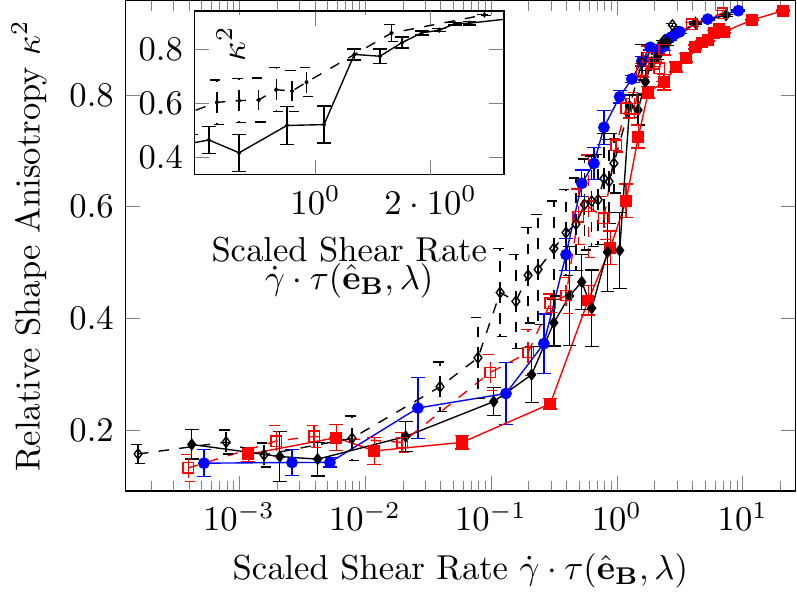}
\\
\tikzFigure{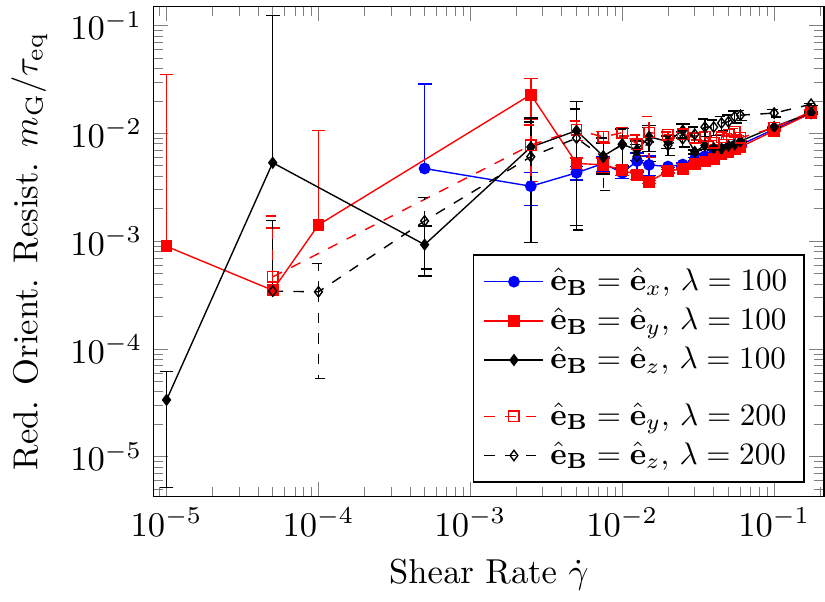}
&
\tikzFigure{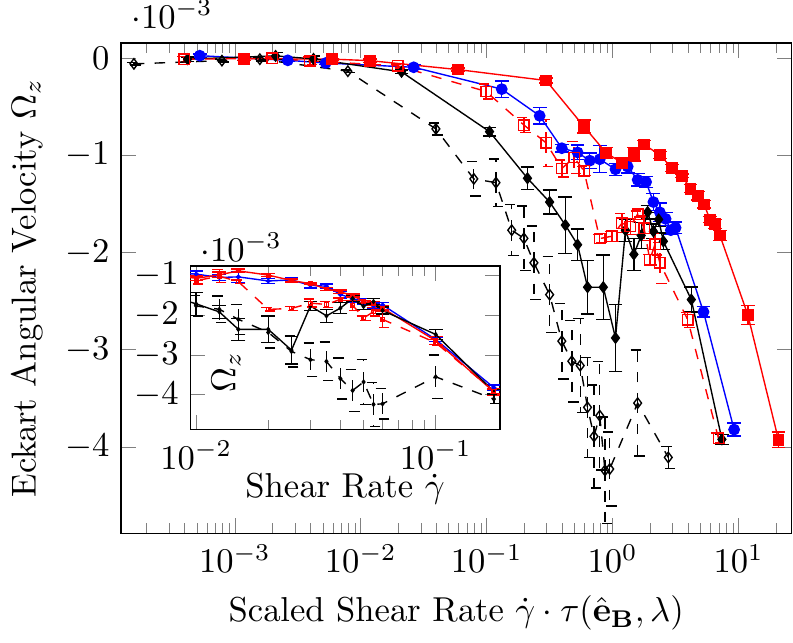}
\end{tabular}
\caption
{
	(color online)
	Specific quantities that characterize the
	magnetically functionalized star polymer as functions of the shear
	rate $\dot \gamma$ for different orientations of
	the external field $\directionB$ and values of its relative strength, $\lambda$
	(as labeled):
	top-left -- number  of magnetic clusters $N_{\textrm{C}}$,
	top-right -- relative shape anisotropy $\kappa^2$,
	inset showing a zoomed-in view for $\directionB = \directionZ$;
	bottom-left -- reduced orientational resistance $m_{\textrm{G}} / \tau_{\textrm{eq}}$,
	and bottom-right -- Eckart angular velocity around $z$-axis $\Omega_z$,
	inset showing data with unscaled shear rates $\dot\gamma$ in a zoomed-in region.
	Scaled shear rates are constructed via an empirically determined $\tau(\directionB,\lambda)$
	such that
	the $N_{\textrm{C}}$ curves collapse to a master curve (cf. inset in top-left figure).
	The legend in the bottom-left panel applies equally to all panels and insets.
	See main text for details.
}
\label{fig:data}
\end{figure*}

\section{Results}
We find that the observed conformational and dynamic properties
can qualitatively be
classified into four categories:
(i)
the mean number of magnetic clusters, $N_{\textrm{C}}$,
which is of particular importance and thus warrants separate treatment,
(ii)
quantities that are largely controlled by $N_{\textrm{C}}$,
(iii)
quantities that are unaffected by the presence of magnetic moments in the model,
and
(iv)
quantities that, \emph{on top} of an $N_{\textrm{C}}$-dependence,
are sensitive to the orientation of the external magnetic field $\directionB$
relative to the shear flow and shear gradient direction.

\subsection{Mean Number of Clusters $N_{\textrm{C}}$}

The main plot in the top-left panel of
\figref{fig:data}
shows the mean number of clusters, $N_{\textrm{C}}$,
as a function of the shear rate $\dot{\gamma}$.
For low $\dot\gamma$-values, the mean cluster count is roughly $2$,
up until a critical shear rate $\dot{\gamma}^\star$ is
reached, which depends on the orientation $\directionB$
and strength (encoded in $\lambda$) of the external magnetic field.
At this $\dot{\gamma}^\star$, shear-induced forces overcome
the attractive magnetic interactions, breaking up
columns of end-monomers (which form along the $\directionB$ direction)
into successively smaller, more stable units
as the shear rate is increased;
to be more specific,
we observe $N_{\textrm{C}} \propto \ln(\dot{\gamma})$.
This critical $\dot{\gamma}^\star$
is largest for $\directionB = \directionZ$
and smallest for $\directionB = \directionY$,
where the magnetic columns are particularly exposed to the shear flow
gradient (cf. \figref{fig:schematic}).
Furthermore, $\dot{\gamma}^\star$, or equivalently,
the robustness of magnetic clusters,
increases with $\lambda$.
The inset shows that, upon scaling the shear rate
with an empirical $\directionB$- and $\lambda$-dependent factor
$\tau(\directionB, \lambda)$,
all curves collapse onto a master curve,
with the scaling chosen such that
$\dot{\gamma}^\star (\directionB, \lambda) \cdot \tau(\directionB, \lambda) \approx 1$.

\subsection{$N_{\textrm{C}}$-Controlled Quantities: Shape Descriptors}

The shape descriptors [cf.\ equation \eqnref{shape_parameters}],
when viewed as functions of the scaled shear rate,
exhibit comparable qualitative behavior for various orientations
$\directionB$ and magnetic interaction strengths $\lambda$.
This is significant in that the shape is largely determined by
the number of magnetic columns formed in a given situation,
but is otherwise relatively unaffected by the details of the magnetic
interaction.

The top-right panel of \figref{fig:data} shows the relative shape anisotropy
$\kappa^2$ as a representative member of this category of quantities.
A value of $\kappa^2$ near $0$ would roughly be indicative of a spherically
symmetric arrangement of the star polymer's beads;
even for low shear rates $\dot\gamma$, this condition is not met, since the
magnetic columns formed by the end-monomers break rotational symmetry,
as they align with the external magnetic field.
For higher shear rates, the polymer is strongly elongated along the flow direction.
Also, note
that there is a sudden increase in $\kappa^2$
at $\dot{\gamma} (\directionB, \lambda) \cdot \tau(\directionB, \lambda) \approx 1$,
i.e. at the critical shear rate $\dot{\gamma}^\star$ where magnetic clusters start breaking
apart, particularly pronounced for $\directionB = \directionZ$ (see panel inset).
Conversely, given the rather well-defined dependence
of $N_{\textrm{C}}$ on ${\bf B}$ and $\dot\gamma$, one can manipulate
the shape and size of the star polymers by tuning the external fields
in their strength and/or relative orientation.

\subsection{Universal Properties: Orientational Resistance}

One can measure the extent of alignment between the flow direction
($\directionX$)
and the major axis of the instantaneous configuration of the star
polymer, i.e. the eigenvector associated with the largest eigenvalue $\Lambda_3^2$ of ${\cal S}$,
and denote the corresponding angle $\chi$;
then, one can define the orientational resistance
$ m_{\textrm{G}} = \dot{\gamma} \tau_{\textrm{eq}} \tan ( 2 \chi ) $,
where $ \tau_{\textrm{eq}} $ is the longest relaxation time of the
star polymer in equilibrium.

The bottom-left panel of \figref{fig:data} shows
$ m_{\textrm{G}} / \tau_{\textrm{eq}} $
as a function of $\dot\gamma$.
For sufficiently large shear rates
($\dot\gamma \gtrsim 10^{-2}$ in inverse MPCD time units),
the orientational resistance follows a power-law
$ m_{\textrm{G}} \propto {\dot{\gamma}}^\mu $
with a characteristic exponent $0.4 < \mu < 0.6$.
This behavior is shared by the majority
of polymeric systems
(each with a corresponding value of $\mu$),
ranging from
linear chains to
block copolymers,
randomly cross-linked single-chain nanoparticles,
dendrimers,
and
non-magnetic star polymers
\cite{Ripoll:2006,Formanek:2018,Huang:2010b,Nikoubashman:2010,Jaramillo-Cano:2018}.
Thus, while the exponent $\mu$ varies with ${\bf B}$,
the characteristic power-law of star polymers is conserved
despite the addition of a magnetic interaction and
the associated introduction of another distinguished axis.

While parts of the literature
predict \cite{Aust:1999}
or report \cite{Ripoll:2006}
$ m_{\textrm{G}} $
approaching a constant plateau for low $\dot\gamma$,
the high fluctuations observed in our data for low shear rates
allow neither confirmation nor dismissal of this claim.

\subsection{$\directionB$-Sensitivity Beyond $N_{\textrm{C}}$: Angular Velocity}

Although the star's shape is largely determined by $N_{\textrm{C}}$,
as discussed above,
the rotational dynamics of the star are peculiar in that
they have an additional dependence on the orientation of $\directionB$:
When considering the angular velocities $\omega_\alpha$ around the
Cartesian axes $\alpha$,
or more appropriately, the Eckart-frame angular velocities $\Omega_\alpha$
--
constructed so as to
remove spurious contributions by vibrational modes to the (apparent)
angular velocity
\cite{Eckart:1935,Louck:1976,Sablic:2017}
--
one finds that there is no net rotation around the $x$ (shear flow direction)
and $y$ (shear gradient direction) axes,
but a significant rotation $\Omega_z \ne 0$
(shear vorticity direction);
this fact additionally and decisively
distinguishes the case
$\directionB = \directionZ$ from
the other ones, even when scaling the shear rates
(cf. \figref{fig:schematic} and bottom-right panel in \figref{fig:data}).

In particular for $\directionB = \directionZ$,
the magnetic interaction parameter $\lambda$ plays no role below the
critical shear rate $\dot{\gamma}^\star$ (see inset),
and as soon as magnetic columns start breaking up, the different curves
approach a common master curve, corresponding to the case of little to
no magnetic clustering. The most pronounced change in (Eckart) angular
velocity occurs, again, at $\dot\gamma = \dot{\gamma}^\star$ (cf. inset)
or
$\dot{\gamma} (\directionB, \lambda) \cdot \tau(\directionB, \lambda) \approx 1$ (cf. main panel),
respectively.

\section{Conclusions and Outlook}

Decorating the arms of star polymers with magnetic particles
opens up a rich, new facet of the phenomenology of polymer
physics.
The resulting magnetically functionalized star polymers are sensitive
to both direction and intensity of an external magnetic field,
as well as to the relative orientation and strength of shear flow.
Said sensitivity manifests in the self-aggregation behavior
of columns of the star's magnetic monomers,
and the stability of the resulting magnetic columns.
This in turn largely determines
size, shape, anisotropy, and dynamic responses,
some aspects of which (e.g. orientational resistance)
behave qualitatively as in the non-magnetic case,
while others (e.g. whole-body rotation) exhibit entirely novel phenomenology.

The tunability of the star conformation, anisotropy, and of the stability of magnetic aggregates
via manipulation of the external magnetic field $\bf B$ allows for new
avenues in which (computer) experiments can be conducted.
For example,
upcoming research will discuss
self-aggregation of magnetic columns
in dense solutions of magnetic stars,
how changes in the external fields
can influence e.g.\ rheology or the formation of
large-scale structures in a given system,
and what types of phase behavior can be observed.
Possible applications might include micro-fluidic devices,
such as micro-mixers with tunable efficiency due solely
to the geometry of flow and magnetic field.

\ack
The authors acknowledge financial support by
the Austrian Science Fund FWF within the SFB ViCoM (F41) and computing
time by the Vienna Scientific Cluster. The authors thank Ronald Blaak
(Clermont-Ferrand) for helpful discussions and Angela Koffler for her help
in creating \figref{fig:schematic}.
D.T.\ and G.K.\ acknowledge financial support by the FWF under Proj.\ No.\ I3846-N36.

\end{document}